# Do Sharpness-based Optimizers Improve Generalization in Medical Image Analysis?

*Mohamed Hassan*, *Aleksandar Vakanski*, *Min Xian*

Computer Science Department, University of Idaho, Moscow, USA

## Abstract

Effective clinical deployment of deep learning models in healthcare demands high generalization performance to ensure accurate diagnosis and treatment planning. In recent years, significant research has focused on improving the generalization of deep learning models by regularizing the sharpness of the loss landscape. Among the optimization approaches that explicitly minimize sharpness, Sharpness-Aware Minimization (SAM) has shown potential in enhancing generalization performance on general domain image datasets. This success has led to the development of several advanced sharpness-based algorithms aimed at addressing the limitations of SAM, such as Adaptive SAM, surrogate-Gap SAM, Weighted SAM, and Curvature Regularized SAM. These sharpness-based optimizers have shown improvements in model generalization compared to conventional stochastic gradient descent optimizers and their variants on general domain image datasets, but they have not been thoroughly evaluated on medical images. This work provides a review of recent sharpness-based methods for improving the generalization of deep learning networks and evaluates the methods performance on medical breast ultrasound images. Our findings indicate that the initial SAM method successfully enhances the generalization of various deep learning models. While Adaptive SAM improves generalization of convolutional neural networks, it fails to do so for vision transformers. Other sharpness-based optimizers, however, do not demonstrate consistent results. The results reveal that, contrary to findings in the non-medical domain, SAM is the only recommended sharpness-based optimizer that consistently improves generalization in medical image analysis, and further research is necessary to refine the variants of SAM to enhance generalization performance in this field.

# 1. Introduction

Deep neural networks (DNNs) have demonstrated success across different tasks in recent years, such as computer vision, natural language processing, and video/speech recognition [1]. However, these networks are typically over-parametrized and are prone to overfitting the training data, resulting in discrepancy in the generalization performance on unseen data [2]. Improving the generalization performance is crucial, as a model's ability to perform well on unseen data is arguably the most important factor determining the algorithm's usability in real-life applications. Generalization poses a greater challenge in medical image analysis, as medical image datasets differ from non-medical image datasets in several ways: they typically contain fewer images, have higher resolution, lower contrast, and a lower signal-to-noise ratio [3]. Additionally, medical images are collected using various devices, protocols, and patient populations, introducing further complexities that hinder generalization on unseen images.

To understand the generalization phenomenon of DNNs, prior studies have developed complexity measures that correlate monotonically with generalization performance. Complexity measures have been proposed based on statistical learning theory bounds and/or empirical studies based on the network size [4, 5], the norm of the model parameters [6, 7], the geometry of the loss landscape [8], and other factors. Jiang et al. [9] evaluated 40 measures using the Network-in-Network (NiN) architecture on the CIFAR-10 and SVHN datasets using Kendall's rank correlation coefficient. Similarly, Dziugaite et al. [10] evaluated 24 complexity measures on CIFAR-10 and SVHN datasets by investigating the causal relation between various measures and generalization using conditional independence tests. Both studies concluded that many norm-based and network size measures performed poorly and stated the potential of PAC-Bayes sharpness-based measures for establishing a connection between the geometry of the loss landscape and generalization. Motivated by this, Vakanski and Xian [3] conducted a correlation experiment akin to Jiang et al. [9] on medical images, assessing the correlation of 25 different complexity measures with generalization using a VGG-like network. The authors reported that PAC-Bayes sharpness measures exhibited the highest positive correlation with generalization on medical images.

The connection between the geometry of the loss landscape and generalization has received a good amount of research as most state-of-the-art overparametrized DNNs, including convolutional neural networks (CNNs) and vision transformers, possess complex landscapes with numerous sharp minima [11]. In such an overparametrized setting, although solutions with sharp minima minimize the training error, many of them do not generalize well [12]. Chaudhari et al. [13] search for flat regions by minimizing the local entropy using an approach suitably called Entropy-SGD. Wen et al. [14] introduced a SmoothOut technique that focuses on smoothening out the sharp minima for higher generalization. Lewkowycz et al. [15] claim that SGD with large learning rates generalizes well since dynamical instability at initialization causes the optimizer to catapult to regions where the maximum eigenvalue of the neural tangent kernel and of the Hessian are smaller. Motivated by previous work, Foret et al. [16] propose Sharpness Aware Minimization (SAM) as a learning algorithm based on the PAC-Bayesian generalization bound that improves model generalization performance on various natural image and language benchmarks by simultaneously minimizing both training loss and sharpness. SAM seeks flat minima by employing adversarial perturbations to maximize the training loss and then minimizing the loss of this perturbed objective through a single update step of a base optimizer, such as the Adam optimizer. However, SAM has limitations in relation to sensitivity to parameter re-scaling and the non-linearity of the loss landscape. These limitations, despite SAM's strong performance, have inspired the development of new sharpness-based optimization algorithms, including Adaptive Sharpness-Aware Minimization (ASAM) [17], Surrogate Gap Guided Sharpness-Aware Minimization (GSAM) [18], Weighted Sharpness-Aware Minimization (WSAM) [19], and Curvature Regularized Sharpness-Aware Minimization (CR-SAM) [20].

Our work was inspired by previous studies that evaluated the generalization performance of SAM and its variants on non-medical benchmark datasets (e.g., CIFAR-10/100 and ImageNet) using CNNs [16, 17, 18, 19, 20, 21] and large language models (LLMs) [22]. These studies demonstrated improvements in model generalization compared to Adam optimizers for all sharpness-based optimizers, including SAM, ASAM, GSAM, WSAM, and CR-SAM. However, prior works have not thoroughly evaluated the effectiveness of sharpness-based models on medical images or with vision transformer models. Additionally, some of the related work focused solely on evaluating generalization performance without validating the geometry of the loss landscape [21, 22].

In this paper, we first survey the relevant recent sharpness-based methods for improving the generalization of neural networks. Next, we assess sharpness-based algorithms on medical breast ultrasound (BUS) images and compare their performance to baseline models using the Adam optimizer. Our objective is to investigate whether SAM approaches enhance the generalization performance of medical images. Toward this goal, we evaluate the performance of classification models, which include two popular CNN-based models, ResNet50 [23] and VGG16 [24], and two Vision Transformer models, ViT [25] and Swin Transformer [26]. We utilize performance metrics derived from the approximation of the Hessian matrix to quantify the connections between the geometry of the loss landscape and generalization. We also report the computational costs of sharpness-based methods to determine their efficiency and scalability for deployment in real-life applications. Our results indicate that SAM is the only sharpness-based optimizer that consistently enhances the generalization performance of all tested models. Additionally, Hessian-based metrics demonstrate a flatter landscape produced by SAM in comparison to the standard Adam optimizer for all tested models. In contrast, ASAM, GSAM, WSAM, and CR-SAM did not exhibit consistent improvements across our experiments. Furthermore, our analysis of computational costs highlights the efficiency of SAM and CR-SAM, which exhibit significantly lower computational overhead compared to ASAM, GSAM, and WSAM, making them more suitable for deployment in real-life applications. Our results reveal that SAM has the highest potential among all sharpness-based optimizers in the domain of medical image analysis.

The main contributions of this work are as follows:

- We overviewed and provided a comparative study of the performance of the most common sharpness-based algorithms: SAM, ASAM, GSAM, WSAM, and CR-SAM.
- We evaluated the generalization performance of sharpness-based algorithms on medical BUS images by utilizing Hessian metrics to investigate changes in the geometry of the loss landscape.
- We compared the generalization performance of sharpness-based algorithms using CNN-based classification models with Vision Transformer models.

## 2. Method

This section first provides an overview of sharpness-based optimizers, including SAM and its variants. Afterward, it presents the metrics used in investigating the geometry of the loss landscape.

**2.1 Sharpness Aware Minimization (SAM)**

The geometry of the loss landscape can influence how well a model generalizes on new datasets. Keskar et al. [8] suggested that models with a flat loss landscape (flat minimum) generalize better than models with a sharp loss landscape (sharp minimum), as the large sensitivity of the training function at a sharp minimizer negatively impacts the trained model's ability to generalize on new data. The generalization gap of DNNs can be formally defined as the difference between the training and test losses. Mathematically, it is defined by:

$$\text{Generalization gap: } L_D(\boldsymbol{w}) - L_S(\boldsymbol{w}), \tag{1}$$

where $\boldsymbol{w}$ is the weight vector, $L_S(\boldsymbol{w})$ is the training loss, $L_D(\boldsymbol{w})$ is the generalization loss. The connection between the generalization gap and sharpness of the loss function within an $\epsilon$-ball is satisfied according to the following definition:

$$\text{Sharpness: } \max_{||\epsilon||_p \leq \rho} L_S(\boldsymbol{w} + \epsilon) - L_S(\boldsymbol{w}), \tag{2}$$

where $\rho$ is the radius of the maximization region which is an $\ell^p$ ball. From the above definition, the sharpness function is the difference between the maximum training loss in the $\ell^p$ ball with a fixed radius $\rho$ and the training loss.

Motivated by this work, Foret et al. [16] propose a learning algorithm called Sharpness Aware Minimization (SAM) that combines the geometry of the loss landscape and a PAC-Bayes norm as a generalization bound. SAM focuses on minimizing the sharpness of the loss landscape during training to seek flat minima by minimizing the following PAC-Bayesian generalization error bound:

$$L_D(\boldsymbol{w}) \leq \max_{||\epsilon||_p \leq \rho} L_S(\boldsymbol{w} + \epsilon) + h\left(\frac{||\boldsymbol{w}||_2^2}{\rho^2}\right), \tag{3}$$

for a strictly increasing function $h$. Because of the monotonicity of $h$, it can be substituted by $\ell^2$ weight decaying regularizer. Hence, SAM can be defined as the following minimax optimization:

$$\min_{\boldsymbol{w}} \max_{||\epsilon||_p \leq \rho} L_S(\boldsymbol{w} + \epsilon) + \frac{\lambda}{2} ||\boldsymbol{w}||_2^2 , \tag{4}$$

where $\lambda$ is a weight decay coefficient.

**2.2 Adaptive Sharpness Aware Minimization (ASAM)**

Kwon et al. [17] claim that some sharpness-based learning methods, including SAM, suffer from sensitivity to model parameter re-scaling, which may weaken the correlation between sharpness and the generalization gap. Sharpness defined in the rigid spherical region with a fixed radius has a weak correlation with the generalization gap because of the non-identifiability of neural networks with ReLU activation functions, whose parameters can be freely re-scaled without affecting its output [27]. The scale-dependency can cause weaker correlation between sharpness and generalization. This can be represented by the following equation:

$$\max_{||\epsilon||_2 \leq \rho} L_S(\boldsymbol{w} + \epsilon) \neq \max_{||\epsilon||_2 \leq \rho} L_S(A\boldsymbol{w} + \epsilon), \tag{5}$$

where A is a scaling operator. Kwon et al. eliminate the vulnerability to weight scaling, by introducing a normalization operator. Considering that $\{T_{\boldsymbol{w}}, \boldsymbol{w} \in \mathbb{R}^k\}$ is a family of invertible linear operators on $\mathbb{R}^k$, if $T_{A\boldsymbol{w}}^{-1} A = T_{\boldsymbol{w}}^{-1}$ for any invertible scaling operator A on $\mathbb{R}^k$ which does not change the loss function, then $T_{\boldsymbol{w}}^{-1}$ is the normalization operator of $\boldsymbol{w}$. Using the normalization operator, Kwon et al. define adaptive sharpness as follows:

$$\max_{||T_{\boldsymbol{w}}^{-1} \epsilon||_p \leq \rho} L_S(\boldsymbol{w} + \epsilon) - L_S(\boldsymbol{w}) \tag{6}$$

The concept of adaptive sharpness is used to formulate Adaptive Sharpness-Aware Minimization (ASAM) as follows:

$$\min_{\boldsymbol{w}} \max_{||T_{\boldsymbol{w}}^{-1} \epsilon||_p \leq \rho} L_S(\boldsymbol{w} + \epsilon) + \frac{\lambda}{2} ||\boldsymbol{w}||_2^2 \quad (7)$$

ASAM induces minimization of the generalization loss without the sensitivity to weight scaling, and due to the negligible calculation cost of $T_{\boldsymbol{w}}^{-1}$, ASAM can be as computationally efficient as SAM.

**2.3 Surrogate Gap Guided Sharpness-Aware Minimization (GSAM)**

Zhuang et al. [18] argue that both sharp and flat minima can exhibit a low perturbed loss, implying that SAM doesn't always prefer flat minima. Motivated by this limitation, Zhuang et al. propose a surrogate gap, a measure equivalent to the dominant eigenvalue of the Hessian matrix at a local minimum when the radius of neighborhood is small. The surrogate gap can hence serve as an equivalent measure of curvature at minima, and it is defined by:

$$h(\boldsymbol{w}) \triangleq L_p(\boldsymbol{w}) - L(\boldsymbol{w}), \quad (8)$$

where $L_p(\boldsymbol{w}) \triangleq \max_{||\delta|| \leq p_t} L(\boldsymbol{w}_t + \delta)$ returns the worst possible loss within a ball of radius $p_t$ centered at $\boldsymbol{w}_t$, and $L(\boldsymbol{w})$ is the loss function with parameter $\boldsymbol{w} \in \mathbb{R}^k$. Using the surrogate gap, Zhuang et al. proposes Surrogate Gap Guided Sharpness Aware Minimization (GSAM) as follows:

$$\min \left( L_p(\boldsymbol{w}), h(\boldsymbol{w}) \right), \quad (9)$$

where $L_p(\boldsymbol{w})$ minimizes the training loss, and $h(\boldsymbol{w})$ minimizes the generalization gap. GSAM consists of two steps; a gradient descent $\nabla L_p(\boldsymbol{w})$ similar to SAM to minimize the perturbed loss $L_p(\boldsymbol{w})$, and an ascent step orthogonal to $\nabla L_p(\boldsymbol{w})$ to minimize the surrogate gap without affecting the perturbed loss.

**2.4 Weighted Sharpness-Aware Minimization (WSAM)**

Yue et al. [19] claim that SAM finds flatter regions but not minima, which could potentially lead to convergence at a point where the loss is still large. This motivated the authors to propose a more general method, called WSAM, by introducing sharpness as a regularization term.

$$L^{WSAM}(\boldsymbol{w}) := L(\boldsymbol{w}) + \frac{\gamma}{1-\gamma} \tilde{L}(\boldsymbol{w}) = \frac{1-2\gamma}{1-\gamma} L(\boldsymbol{w}) + \frac{\gamma}{1-\gamma} L^{SAM}(\boldsymbol{w}) \quad (10)$$

In the above equation, $L(w)$ is the loss function, $\tilde{L}(w)$ approximates the dominant eigenvalue of the Hessian at local minima, and $\frac{\gamma}{1-\gamma}\tilde{L}(w)$ is the regularized weighted sharpness term. The hyperparameter $\gamma \in [0, 1)$ controls the weight of the sharpness to direct the loss trajectory to find either flatter or lower minima. When $\gamma = 0$, the loss term $L^{WSAM}(w)$ degenerates to a vanilla loss, when $\gamma = \frac{1}{2}$, $L^{WSAM}(w)$ is equivalent to the original SAM algorithm (i.e., $L^{SAM}(w)$), and when $\gamma > \frac{1}{2}$, the term $L^{WSAM}(w)$ gives more weights to the sharpness, so that it finds the point which has smaller curvature rather than smaller loss.

To control the regularization term, Yue et al. use a weight decouple technique inspired by Loshchilov and Hutter [28] so that it reflects the sharpness of the current step without additional information, where $\tilde{L}(w)$ is not integrated into the base optimizer to calculate the gradients and update weights, but it is calculated independently. The authors claim that WSAM is more efficient than SAM as it can achieve different minima by choosing different values of $\gamma$.

### 2.5 Curvature Regularized Sharpness-Aware Minimization (CR-SAM)

Wu et al. [20] posit that because SAM aims to improve generalization by minimizing worst-case loss using one-step gradient ascent as an approximation, such one-step gradient ascent approach becomes less effective as the training progresses due to the non-linearity of the loss landscape. On the other hand, using a multi-step ascent gradient approach will cause higher training costs. To tackle this issue, Wu et al. propose Curvature Regularized SAM (CR-SAM) by integrating a normalized Hessian trace $C(w)$ that measures the curvature of the loss landscape.

$$C(w) = \frac{TR(\nabla^2 L_S(w))}{||\nabla L_s(w)||_2} \tag{11}$$

The issue with the normalized Hessian trace is that a direct minimization of $C(w)$ would lead to an increase in the gradient norm $||\nabla L_s(w)||_2$, which could adversely affect generalization. Therefore, Wu et al. optimize $TR(\nabla^2 L_S(w))$ and $||\nabla L_s(w)||_2$ separately as shown below:

$$R_c(w) = \alpha log TR(\nabla^2 L_S(w)) + \beta \log ||\nabla L_s(w)||_2 \tag{12}$$

In the above formula, $\alpha > \beta > 0$, so that the numerator of $C(w)$ is penalized more than the denominator. The term $R_c(w)$ is a combination of normalized Hessian trace with gradient norm penalty regularizer, and is used as a regularization term in SAM as shown below:

$$L^{CR-SAM}(w) := L^{SAM}(w) + R_c(w) \tag{13}$$

Since computing the Hessian trace as in $R_c(w)$ for large matrices in over-parameterized DNNs can be computationally intensive, Wu et al. proposes an approximation based on finite difference (FD), which reduces the complexity in a computation parallelizable way. Wu et al. state that the parallelism in CR-SAM optimizes the worst-case and best-case perturbations within the parameter space ($\rho$-bounded neighborhood) promoting a smoother, flatter loss function with same training speed as SAM.

**2.6 Hessian-based Metrics**

Model analysis by using metrics derived from the Hessian matrix is widely utilized in scientific computing. To address the loss landscape curvature information, we use the PyHessian [29] framework, which directly analyzes Hessian (second-derivative) information with respect to the model parameters. Considering a deep learning network with *m* parameters, the gradient of the loss with respect to model parameters is a vector:

$$\frac{\partial L}{\partial \theta} = g_\theta \in \mathbb{R}^m , \qquad (14)$$

and the second derivative of the loss is the matrix:

$$H = \frac{\partial^2 L}{\partial \theta^2} = \frac{\partial g_\theta}{\partial \theta} \in \mathbb{R}^{m \times m} \qquad (15)$$

Given that deep learning networks typically have millions or billions of parameters, forming the Hessian matrix explicitly becomes computationally unfeasible. Instead, properties of the Hessian spectrum are often computed without explicitly forming the matrix. This is achieved by using an oracle to compute the multiplication of the Hessian to a random vector $v$, as shown below:

$$\frac{\partial g_\theta^T v}{\partial \theta} = \frac{\partial g_\theta^T}{\partial \theta} v + g_\theta^T \frac{dv}{d\theta} = \frac{\partial g_\theta^T}{\partial \theta} = Hv, \qquad (16)$$

where the cost of this Hessian matrix-vector multiply (referred to as a Hessian matvec) is the same as one gradient backpropagation. Having this Hessian matvec, the top Hessian eigenvalue is computed using power iteration [30] to reveal the principal curvatures of the loss landscape, indicating the directions of greatest descent or ascent. Positive eigenvalues suggest an ascent toward a local minimum, while negative eigenvalues indicate a descent from a local minimum. Smaller eigenvalues (closer to 0) generally correspond to flatter regions in the loss landscape, which are associated with better generalization performance.

To obtain additional information on the geometry of the loss landscape, the trace of the Hessian can be computed using Hutchinson's method [31] by sampling the random vector $v$ from a Gaussian distribution with mean 0 and variance 1, resulting in the following identity:

$$Tr(H) = Tr(HI) = Tr(H\mathbb{E}[vv^T]) = \mathbb{E}[Tr(Hvv^T)] = \mathbb{E}[v^T H v], \qquad (17)$$

where $I$ is an identity matrix of appropriate size. The trace of the matrix $Tr(H)$ is estimated by computing the expectation of drawing multiple random samples $\mathbb{E}[v^T H v]$, where $v^T H v$ is a dot product between the Hessian matvec and the original vector $v$.

In addition to the top Hessian eigenvalue and the Hessian trace, we compute the median, mean and standard deviation of the Hessian eigenvalues by sampling 100 different weight perturbations for every batch in the train set. This information offers a more comprehensive understanding of the curvature of the loss landscape. While the top Hessian eigenvalue and the Hessian trace provide valuable information about the largest curvature of the loss landscape, they may not fully represent the overall shape and distribution of the curvatures. A loss landscape with low mean, median, and standard deviation values typically indicate smoother optimization paths.

## 3. Experimental Results

### 3.1 Breast Ultrasound Dataset

Breast cancer ranks as the most frequently diagnosed cancer among women and remains one of the leading causes of cancer-related deaths worldwide [32]. The primary goal in combating breast cancer is to reduce its mortality rate by identifying signs and symptoms at an early stage. Deep learning models have been applied to breast cancer diagnosis via mammography, ultrasonography, and magnetic resonance imaging [33]. However, medical images often suffer from poor image quality due to factors such as speckle noise, low contrast, weak boundary definition, and variations in tumor size and echo strength across patients [34,35]. These challenges contribute to the limited generalization performance of deep learning models in this domain. In this paper, we use a BUS dataset, GDPH&SYSUCC, which is the largest open dataset currently in this field [36]. This dataset is collected from two medical centers: the Department of Ultrasound, Guangdong Provincial People's Hospital (GDPH) and the Department of Ultrasound, Sun Yat-sen University Cancer Center (SYSUCC). The images were acquired with the following equipment: Mindray DC-80 from China, Hitachi Ascendus and Toshiba Aplio 500 from Japan, and Supersonic Aixplorer from France. It consists of 886 benign and 1,519 malignant images, for a total of 2,405 BUS images. BUS images include four layers: fat layer, gland layer, muscle layer, and thorax layer (as shown in Figure 1), where malignant tumors start from the gland layer and tend to invade into deeper layers vertically, while benign tumors start at the gland layer and likely to extend horizontally within the gland.

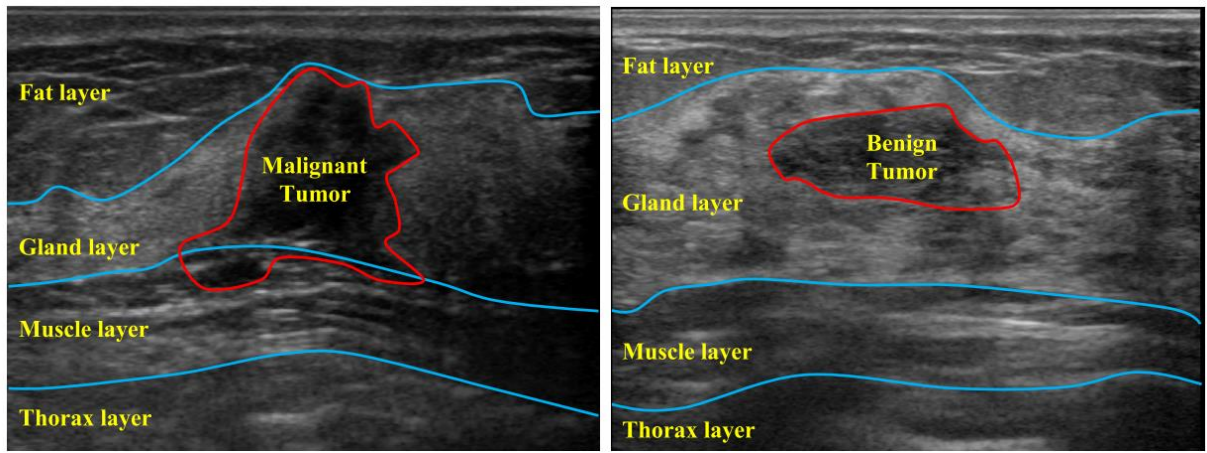

Figure 1: GDPH&SYSUCC Dataset sample Malignant image (left) and Benign image (right).

**3.2 Experiments**

**Experiment 1: Baseline Measurement**

To establish a baseline measurement for comparing the sharpness-based optimizers, we used subsets of the train (80%) and test (20%) images, an Adam optimizer with a learning rate of $10^{-4}$, and a batch size of 16. We trained two Convolution Neural Network models, ResNet50 and VGG16, and two Vision Transformer models, ViT and Swin transformer, until achieving 100% training accuracy, and we recorded the corresponding testing accuracy and training speed using Falcon GPU. Additionally, the top Hessian eigenvalue, trace, median, mean, and standard deviation are all utilized to evaluate the sharpness of the loss landscape for each sharpness-based optimizer, where smaller values of these metrics (closer to 0) indicate a flatter loss landscape.

**Experiment 2: Baseline Measurement vs Sharpness-based optimizers with ResNet50**

ResNet50 [23] is a 50-layer convolutional neural network (comprising 48 convolutional layers, one MaxPool layer, and one average pool layer) that forms networks by stacking residual blocks. This experiment compares the baseline performance of ResNet50 using Adam optimizer against sharpness-based optimizers to determine their effect on generalization, training speed, and the loss landscape, while keeping the hyperparameters the same as in Experiment 1.

From Table 1, VGG16 demonstrates the highest generalization ability with a test accuracy of 84% on unseen data, followed by Swin Transformer at 80.5%, ResNet50 at 80.2%, and ViT at 80%. ViT shows the flattest loss landscape with the lowest Hessian eigenvalue components, and ResNet50 shows the sharpest loss landscape. This is consistent with previous studies as the loss function of ViT is non-convex, while that of ResNet50 is strongly convex [37]. Additionally, ViT achieves the fastest training speed, taking only 2 minutes, whereas Swin Transformer is the most computationally expensive, requiring 21 minutes.

| Model | Test Accuracy (%) | Training time (min) | Top Hessian Eigenvalue | Hessian Median | Hessian Mean | Hessian SD | Hessian Trace |
|---|---|---|---|---|---|---|---|
| **ResNet50** | 80.2 | 12 | 10631.19 | 17088.91 | 31078.76 | 34360.13 | 18202.65 |
| **VGG16** | **84** | 14 | 32.69 | 98.06 | 152.91 | **168.26** | 80.26 |
| **ViT** | 80 | **2** | **19.42** | **21.47** | **127.58** | 390.50 | **36.14** |
| **Swin** | 80.5 | 21 | 206.82 | 130.64 | 136.59 | 341.54 | 466.98 |

Table 1: Baseline Performance ResNet50, VGG16, ViT, and Swin Transformer models using Adam optimizer.

| Model | Test Accuracy (%) | Training time (min) | Top Hessian Eigenvalue | Hessian Median | Hessian Mean | Hessian SD | Hessian Trace |
|---|---|---|---|---|---|---|---|
| **Adam** | 80.2 | 12 | 10631.19 | 17088.91 | 31078.76 | 34360.13 | 18202.65 |
| **SAM** | **84** | 38 | 384.33 | 230.23 | 463.76 | 692.36 | 712.93 |
| **ASAM** | **84.2** | 420 | **7.46** | **7.87** | **8.67** | **3.41** | **34.08** |
| **GSAM** | **84.2** | 1440 | 260.52 | 88.84 | 286.27 | 654.28 | 314.97 |
| **WSAM** | 79.4 | 1450 | 1508.93 | 214.96 | 2419.83 | 5865.05 | 1651.63 |
| **CRSAM** | 76.3 | **22** | 16695.02 | 13893.18 | 15517.05 | 8345.49 | 27176.58 |

Table 2: Baseline Performance of ResNet50 vs the most common sharpness-based optimizers

The results from Table 2 indicate that SAM, ASAM, GSAM, and WSAM all result in a flatter loss landscape compared to the baseline Adam optimizer, whereas CRSAM shows a sharper loss landscape with high Hessian eigenvalue components. SAM, ASAM, and GSAM demonstrate higher generalization ability, with test accuracies higher than the baseline accuracy (80.2%), while WSAM and CRSAM show lower test accuracies. ASAM and GSAM both achieve the highest generalization ability with a test accuracy of 84.2%, followed closely by SAM with a test accuracy of 84.0%. However, SAM exhibits the fastest training speed at 38 minutes, followed by ASAM at 420 minutes and GSAM at 1440 minutes.

**Experiment 3: Baseline Measurement vs Sharpness-based optimizers with VGG16**

VGG16 [24] is a deep convolutional neural network known for its simplicity and uniform architecture with just 16 layers (13 convolutional layers and 3 fully connected layers). This experiment compares the baseline performance of VGG16 using Adam optimizer against sharpness-based optimizers. The testing environment is similar to Experiment 2.

| Model | Test Accuracy (%) | Training time (min) | Top Hessian Eigenvalue | Hessian Median | Hessian Mean | Hessian SD | Hessian Trace |
|---|---|---|---|---|---|---|---|
| **Adam** | 84 | 14 | 32.69 | 98.06 | 152.91 | 168.26 | 80.26 |
| **SAM** | **87.5** | 24 | **0.01** | **0.03** | **0.47** | 1.46 | **0.02** |
| **ASAM** | **89.2** | 840 | 5.32 | 5.24 | 5.55 | **0.92** | 18.46 |
| **GSAM** | 78.2 | 1200 | 19.49 | 19.67 | 9.57 | 21.29 | -192.55 |
| **WSAM** | 84 | 1460 | 86.42 | 93.91 | 98.94 | 38.01 | 65.32 |
| **CRSAM** | **86.3** | **16** | 3935.38 | 478.29 | 245.04 | 982.65 | 3790.39 |

Table 3: Baseline Performance of VGG16 vs the most common sharpness-based optimizers

The results in Table 3 show that SAM, ASAM, WSAM and GSAM result in a flatter loss landscape than the baseline Adam optimizer, while CR-SAM results in a sharper loss landscape, which is consistent with our findings in Experiment 2. Also consistent with Experiment 2, ASAM and SAM exhibit the highest generalization ability with test accuracies of 89.2% and 87.5%, respectively. CR-SAM also demonstrates better generalization ability than the baseline, with a test accuracy of 86.3%, while WSAM shows equal generalization ability at 84%, and GSAM exhibits the worst generalization ability at 78.2%. SAM and CR-SAM are significantly faster than the other sharpness-based optimizers, with training speeds of 24 minutes and 16 minutes, respectively, while ASAM is the third fastest with a training speed of 840 minutes.

Experiments 2 and 3 indicate that SAM and ASAM have the best overall performance compared to the Adam optimizer and other flatness methods when using ResNet50 and VGG16.

**Experiment 4: Baseline Measurement vs Sharpness-based optimizers with ViT**

Vision Transformers (ViT) [25] is an architecture that uses self-attention mechanisms to process images using transformer blocks, where each block consists of two sub-layers; a multi-head self-attention layer and a feed-forward layer. This experiment compares the baseline performance of ViT using Adam optimizer against sharpness-based optimizers to evaluate if they cause any effect on generalization, training speed and the loss landscape, while keeping the hyperparameters the same as the previous experiments.

| **Model** | **Test Accuracy (%)** | **Training time (min)** | **Top Hessian Eigenvalue** | **Hessian Median** | **Hessian Mean** | **Hessian SD** | **Hessian Trace** |
|---|---|---|---|---|---|---|---|
| **Adam** | 80 | 2 | 19.42 | 21.47 | 127.58 | 390.50 | 36.14 |
| **SAM** | **82.5** | 4 | **0.65** | **12.48** | **11.08** | 11.37 | 4.83 |
| **ASAM** | 69.6 | 53 | 82.90 | 40.47 | 43.69 | 39.36 | 1104.52 |
| **GSAM** | 77.3 | 56 | 17.49 | **12.48** | 12.26 | 11.37 | -165.05 |
| **WSAM** | 73.6 | 32 | 19.38 | 22.67 | 22.70 | **5.94** | 5.87 |
| **CRSAM** | 79 | **3** | 0.79 | 16.77 | 271.53 | 646.09 | **2.84** |

Table 4: Baseline Performance of ViT vs the most common sharpness-based optimizers

The results from Table 4, show that SAM, GSAM, and CRSAM result in a flatter loss landscape than the regular Adam optimizer, as evidenced by their lower Hessian eigenvalue components, while ASAM and WSAM show a sharper loss landscape. However, the only sharpness-based optimizer that shows higher generalization ability is SAM as it results in 82.5% test accuracy, while the rest result in a test accuracy less than the baseline accuracy (80%). SAM also shows high training speed with just 4 minutes.

**Experiment 5: Baseline Measurement vs Sharpness-based optimizers with Swin Transformer**

Swin Transformer [26] is a hierarchical visual transformer with an efficient shift-window partitioning scheme for computing self-attention. In this experiment, we compare the baseline performance of the Swin Transformer using the Adam optimizer to that of the sharpness-based optimizers. We evaluate generalization, training speed, and the geometry of the loss landscape, ensuring that the hyperparameters remain consistent with those used in previous experiments.

| Model | Test Accuracy ( % ) | Training time (min) | Top Hessian Eigenvalue | Hessian Median | Hessian Mean | Hessian SD | Hessian Trace |
|---|---|---|---|---|---|---|---|
| **Adam** | 80.5 | 21 | 206.82 | 130.64 | 136.59 | 341.54 | 466.98 |
| **SAM** | **82.3** | 48 | **18.83** | **17.27** | **20.31** | 88.63 | **93.13** |
| **ASAM** | 60.7 | 650 | 359.04 | 359.73 | 359.72 | 0.83 | 440.36 |
| **GSAM** | 60.7 | 720 | 300.37 | 300.39 | 300.37 | **0.07** | 4196.10 |
| **WSAM** | 66.3 | 380 | 680.60 | 684.70 | 685.27 | 27.19 | 1729.92 |
| **CRSAM** | 69.4 | **45** | 135.03 | 143.45 | 143.46 | 38.58 | 111.71 |

Table 5: Baseline Performance of ViT vs the most common sharpness-based optimizers

The results from Table 5 show that only SAM and CR-SAM result in a flatter loss landscape compared to the baseline Adam optimizer, as indicated by their lower Hessian eigenvalue components, while the other methods exhibit a sharper loss landscape. Consistent with the results in Experiment 4, only SAM demonstrates higher generalization performance than the baseline Adam optimizer, with a test accuracy of 82.3%, whereas the other sharpness-based optimizers result in test accuracies below 80.5%.

# 4. Discussion

In this work, we evaluate the generalization performance of the most common sharpness-based optimizers for a medical image classification task, comparing them to the best baseline generalization performance achieved with the Adam optimizer. Toward this goal, we conducted experiments using two CNN-based classification models (ResNet50 and VGG16) and two Vision Transformer models (ViT and Swin Transformer). Our results show that SAM consistently achieved higher generalization performance than the Adam optimizer across all tested models. ASAM demonstrated improved generalization performance for CNN-based models but performed poorly for Vision Transformers. Additionally, GSAM improved generalization performance only for ResNet50, while CR-SAM showed improvements only for VGG-16. WSAM, however, failed to enhance generalization performance for any of the tested models.

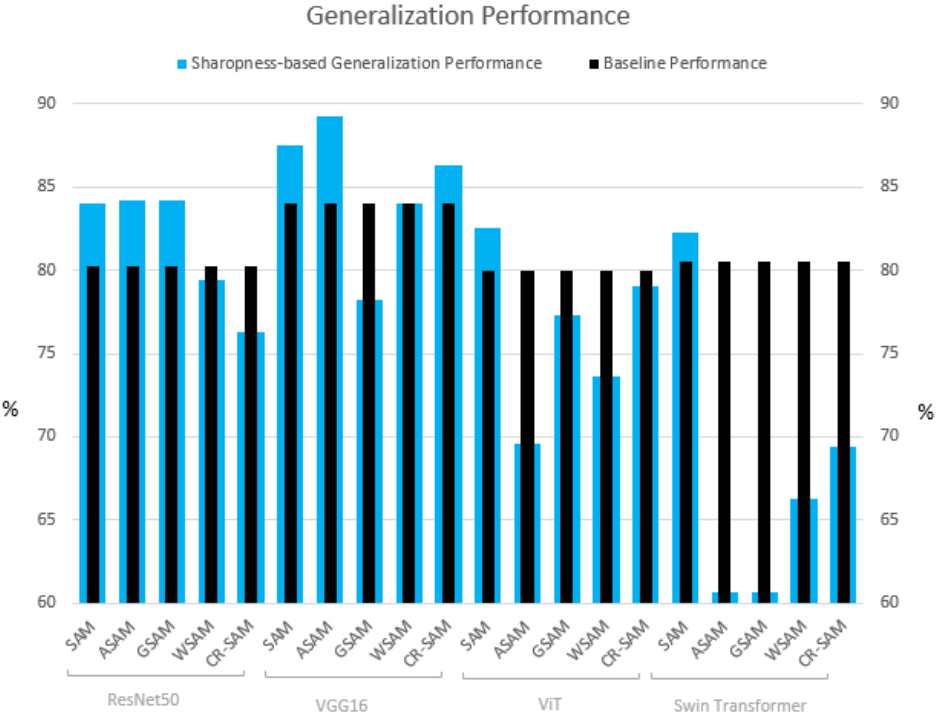

Figure 2: Generalization Performance of sharpness-based optimizers against the best baseline performance of Adam optimizer.

Our results, as illustrated in Figure 2, show that SAM is the only recommended sharpness-based optimizer for improving generalization performance on BUS images. Hessian calculations validate that the loss landscape produced by SAM is flatter than that resulting from the standard Adam optimizer, supporting the hypothesis that flatter minima leads to better generalization.

In future work, we plan to generate plots to visualize the geometry of the loss landscape for a better understanding of the connection between sharpness and generalization. Additionally, we aim to combine SAM with other regularizers and improve SAM's variants to further enhance generalization performance in medical image analysis.

# 5. Conclusion

This paper provides a survey on the most common sharpness-based algorithms for improving the generalization of neural networks. To the best of our knowledge, this paper is the first to evaluate sharpness-based optimizers for their usability in medical image analysis, specifically for CNN-based classification models and Vision Transformers. Our results indicate that Sharpness-Aware Minimization (SAM) is the only method that consistently enhances the generalization ability of all tested models. Performance comparisons based on Hessian calculations demonstrate that the loss landscape produced by SAM is flatter than that resulting from the standard Adam optimizer for all tested models. In contrast, ASAM, GSAM, WSAM, and CR-SAM did not exhibit consistent improvements across our experiments. These findings suggest that more research is necessary to refine SAM and its variants to further enhance generalization in medical image analysis.